\documentclass[aps,prc,showpacs,twocolumn,superscriptaddress]{revtex4}
\usepackage{multirow}
\usepackage{graphicx}
\usepackage{amsmath}

\usepackage{color}

\begin{document}
\title{
\begin{flushright}
{
}
\end{flushright}
Higher Moments of Net-Baryon Distribution as Probes of QCD Critical
Point}

\author{ Y.~Zhou$^{1,2 *}$, S.~S.~Shi$^{1,2}$, K.~Xiao$^{1,2}$, K.~J.~Wu$^{1,2}$ and F~.Liu}

\address{Institute Of Particle Phyisics, HuaZhong Normal
University(CCNU), Wuhan 430079, People's Republic of China}

\address{Key Laboratory of Quark $\&$ Lepton Physics, HuaZhong Normal
University(CCNU), Ministy of Education, \\Wuhan 430079, People's
Republic of China
\\
$^{*}$zhouy@iopp.ccnu.edu.cn}

\date{\today}

\begin{abstract}
It is crucially important to find an observable which is independent on the 
acceptance and late collision process, in order to search for the possible
Critical Point predicted by QCD. By utilizing A Multi-Phase
Transport (AMPT) model and Ultra Relativistic Quantum Molecular
Dynamics (UrQMD) model, we study the centrality and evolution time
dependence of higher moments of net-baryon distribution in Au + Au
collisions at $\sqrt{s_{NN}}=17.3$ GeV. The results suggest that
Kurtosis and Skewness are less sensitive to the acceptance effect
and late collision process. Thus, they should be good observables
providing the information of the early stage of heavy ion collision.
In addition, our study shows that the Kurtosis times $\sigma^{2}$ of
net-proton distribution are quite different to that of net-baryon
when collisions energy is lower than $\sqrt{s_{NN}}$ = 20 GeV, the
Monte Calor calculations on Kurtosis$\cdot\sigma^{2}$ have a
deviation from the theoretical predictions.
\end{abstract}


\maketitle
\clearpage
\section{Introduction}
In QCD phase diagram, the first order phase transition boundary
separates the two phases: the hadron gas and the quark gluon
plasma(QGP). The Critical Point(CP) which locates at the end of the
transition boundary, is a distinct singular feature of QCD phase
diagram\cite{ZF-JHEP}. To map the components of the QCD phase
diagram is one of the main goals of heavy-ion physics, and searching
for the critical point has been addressed in both theoretical and
experimental
studies\cite{ZF-NATURE,MAS-PRL,RVG-PRD,STAR-Note,TS-PoS,JMH-CBM},
all these works help us to gather information of the singularities
near the CP. If the CP existence around
$\mu_{B}\sim$300MeV\cite{RVG-PRD} it can be studied by heavy ion
experiments, like BNL Relativistic Heavy Ion Collider (RHIC) with
its Beam Energy Scan (BES) program\cite{STAR-Note}, CERN Super
Proton Synchrotron (SPS)\cite{TS-PoS} and the future GSI Facility
for Antiproton and Ion Research (FAIR)\cite{JMH-CBM}.

The fluctuations of conserved quantities, like net baryon number,
electric charge and strangeness, are considered to be sensitive
indicators for the structure of created system~\cite{SJ-FL}. The
characteristic feature of a CP is the diverge of the fluctuations.
Most proposed fluctuation of observables are variations of
$2^{\mathrm {rd}}$ order moments of the distribution, such as
particle ratio~\cite{SJ-FL}, charged dynamical~\cite{STAR-Dyn} and
$\Phi$-measure~\cite{SM-PRC}. It has been proved that all these
observables are proportional to approximately
$\xi^{2}$~\cite{MAS-PRD}, where $\xi$ is the correlation length
which will diverges at the CP. Theoretical calculations predict that
$\xi\approx2-3$ fm for heavy ion collision~\cite{BB-CL}, thus, it is
extremely difficult to measure in experiments. However, the recent
results shown that higher moments of conserved quantities
distribution are more sensitive to CP due to their strong dependence
on $\xi$, the $3^{\mathrm {rd}}$ order moment
Skewness$\sim\xi^{4.5}$ while the $4^{\mathrm {th}}$ order moment
Kurtosis$\sim\xi^{7}$~\cite{MAS-PRL}. Also the product
Kurtosis$\cdot\sigma^{2}$ (called $K^{\mathrm {eff}}$ in
reference~\cite{TS-arXiv-Keff}) which is equal to the ratio of
$4^{\mathrm {th}}$ order to $2^{\mathrm {nd}}$ order
susceptibilities shows a large deviation from unity near the CP by
lattice calculations and QCD based model studies~\cite{MC-PRD}.

Studying the event-by-event fluctuations~\cite{GB-EbE,HH-EbE} is an
effective way to address fluctuations of a system created in a heavy
ion collision~\cite{VK-EbE}. In this paper, we will present the
study of higher moments as a function of evolution time, and we will
discuss the acceptance effect on higher moments of net baryon
distribution.

\section{\bf Monte Carlo models}
Monte Carlo event generators, AMPT and UrQMD, have been used in this
study. A Multi-Phase Transport(AMPT) model is made up by four main
parts: the initial conditions, partonic interactions, hadronization
and hadronic rescattering. The initial conditions, which include the
spatial and momentum distributions of minijet partons and soft
string excitations, are obtained from the Heavy Ion Jet Interaction
Generator(HIJING) model~\cite{HIJING}. Scatterings between partons
are modeled by Zhang's Parton Cascade(ZPC)~\cite{ZB}, which
presently includes only two-body scatterings with cross sections
obtained from the pQCD with screening masses. In the default AMPT
model(abbr. ``AMPT Default'')~\cite{ZWL}, partons are recombined
with their parent strings when they stop interacting, and the
resulting strings are converted to hadrons using the Lund string
fragmentation model~\cite{BA,TS}. In the AMPT model with string
melting(abbr. ``AMPT StringMelting'')~\cite{ZWL-JPG}, The transition
from the partonic matter to the hadronic matter is achieved by a
simple coalescence model, which combines two quarks into mesons and
three quarks into baryons~\cite{LWC-PLB}. The authors of AMPT model
use hadronic cascade, which is based on the A Relativistic
Transport(ART) model~\cite{BAL-PRC}, to describe the dynamics of the
subsequent hadronic matter. Final results from the AMPT model are
obtained after hadronic interactions are terminated at a cutoff time
$t_{\mathrm {cut}}$. When setting the time-step(in $\mathrm {fm}/c$)
for hadron cascade to 0.2 (default value), the termination time of
hadron cascade $t_{cut}=0.2\times \mathrm {NTMAX}$. Here $\mathrm
{NTMAX}$ is the number of time-steps, the default setting is
$\mathrm {NTMAX}$. When the time-step is fixed, larger $\mathrm
{NTMAX}$ means longer time of hadron rescatterings. Note $\mathrm
{NTMAX}$ lets all resonance rapidly decay(less than $0.6 \mathrm
{fm}/c$) and turns off hadronic interactions effectively.
In this paper, we measure the fluctuation which belongs to different
processes: scenario (a)``parton'' is a process between ZPC and quark
coalescence, it still locates in the partonic phase, scenario
(b)``w/o ART'' is the time that hadronization(or quark coalescence)
has finished, it is in hadronic phase but no hadron cascade happens,
scenario (c)and (d)are originated from different termination
time $t_{\mathrm {cut}}$, by setting different $\mathrm {NTMAX}$ value we can control
the hadronic process in the simulated heavy ion collisions .
It helps us to understand the hadronic effect on the observable.
In order to investigate different effects in time evolution,
we will study the fluctuation of higher moments in each process.
Comparing the measured value from different processes,
we can study whether the fluctuation of the higher moment can provide the information
from the early stage of the collisions.

The Ultra Relativistic Quantum Molecular Dynamics (UrQMD)
model~\cite{UrQMD} is also been used in this paper. It is a
microscopic transport theory based on the covariant propagation of
all hadrons on classical trajectories in combination with stochastic
binary scatterings, color string formation and resonance decay. The
UrQMD model represents a Monte Carlo solution of a large set of
coupled partial integro-differential equations for the time
evolution of the various phase space densities $f_{i}(x,p)$ of
different species of particles. In the input file, one can control
the time to propagate and output time-interval (in fm/c), in this
paper we set the time to propagate is 40
fm/c and the time-interval is 2 $\mathrm {fm}/c$. More detail
descriptions can be found in Ref.~\cite{UrQMD}.

\section{Results and Discussion}
\begin{figure}
\resizebox{!}{80mm}{\includegraphics{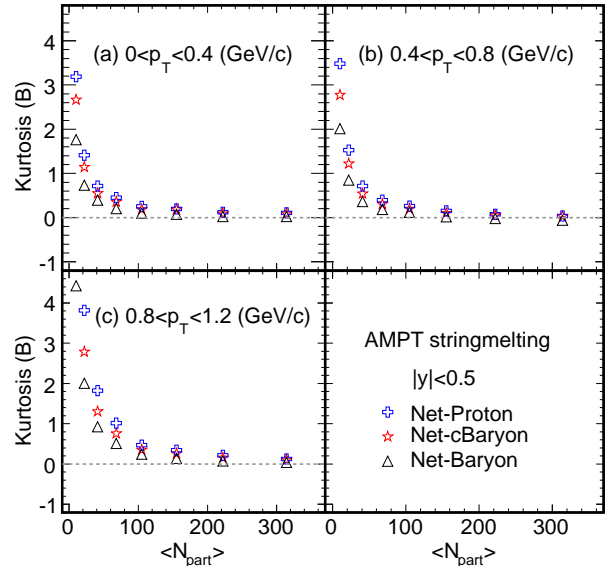}}
\caption{\label{fig:epsart}(Color online) The centrality dependence
of Kurtosis of net-proton, net-cBaryon, net-Baryon distribution in
different transverse momentum windows at $\sqrt{s_{NN}}=17.3$ GeV by
AMPT StringMelting.}
\end{figure}

\begin{figure}
\resizebox{!}{60mm}{\includegraphics{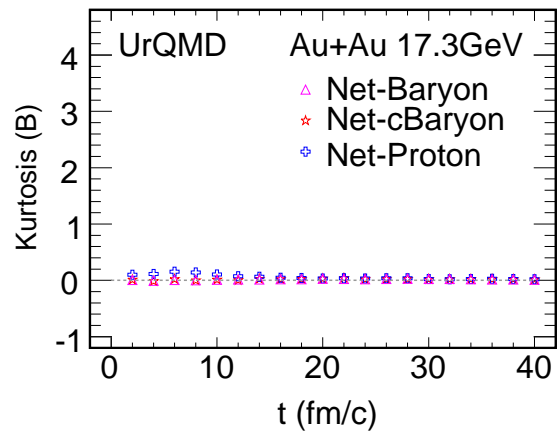}}
\caption{\label{fig:epsart} (Color online) The time dependence of
Kurtosis of net-proton, net-cBaryon, net-Baryon distribution with
b=2, $|y|<0.5$ and $0<p_{T}<1$ ($\mathrm {GeV}/c$) by UrQMD model.}
\end{figure}

In experiments, neutrons can not be detected, and the reconstruction
efficiency is relatively low for strange hadrons, especially for
multi-strange hadrons, such as $\Xi$ and $\Omega$~\cite{SSS-PRC}.
Fortunately, theoretical calculation suggests that the proton number
is a meaningful observable for the purpose of detecting the CP in
heavy ion experiments~\cite{YH-PRL}, its fluctuation completely reflect the
singularity of the baryon number susceptibility. Thus,
only if the measurements from net baryon and net proton distribution
are similar, we can searching for the CP by measuring the
fluctuation of various moments of net proton distribution.

\label{sect_discuss}
\begin{figure*}[ht]
\hskip -2.0cm
\includegraphics[width=0.95\textwidth]{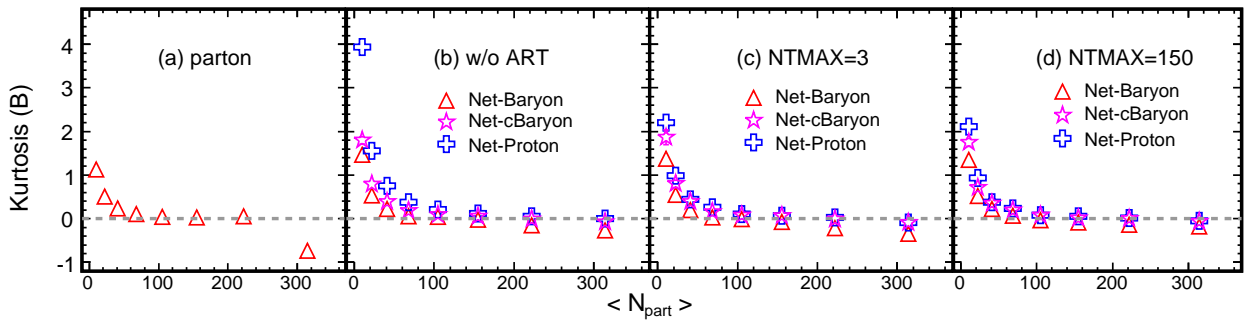}
\caption{ (Color online) The centrality dependence of Kurtosis of
net-proton, net-cBaryon and net-baryon distribution in time
evolution at $\sqrt{s_{NN}}=17.3$GeV within AMPT StringMelting.}
\end{figure*}

Observables that can reflect real dynamics and have little influence
on the finite acceptance, are worth exploring from the theoretical
point of view~\cite{MB-PRC,SM-PLB}. Due to the limited acceptances
in experiments, it is necessary to study the acceptance effect by
Monte Calor model. In the previous work~\cite{ZHOUY-MPE}, we
presented that the Skewness and Kurtosis have little influence on
the size of rapidity windows and transverse momentum regions, they
could be good candidate for searching the CP in experiments.

Firstly, we study the transverse momentum window cut effect
on higher moments. Figure 1 shows the centrality dependence of
Kurtosis of net-proton, net-cBaryon, net-Baryon distribution  at
$\sqrt{s_{NN}}=17.3$ GeV by AMPT StringMelting. The transverse momentum window are
0$<p_{T}<$0.4 ($\mathrm {GeV}/c$), 0.4$<p_{T}<$0.8 ($\mathrm
{GeV}/c$) and 0.8$<p_{T}<$1.2 ($\mathrm {GeV}/c$), respectively.
The only difference between net-Baryon and net-cBaryon is that
net-cBaryon excludes neutrons. As we presented
before~\cite{ZHOUY-MPE}, the Kurtosis has a decreasing trend from
peripheral to central collision. The value of Kurtosis approaches to zero in the central collisions,
it means that these distributions are more similar to Gaussian
distribution.
Based on discussions in Ref.~\cite{LXF-arXiv}, Kurtosis can be scaled by $\langle N_\mathrm{part} \rangle$,
if Kurtosis deviate from this $\langle N_\mathrm{part} \rangle$ scaling curve and exhibit a non-monotonic behavior, it would indicate the new physics which is related to CP.

As shown in each panel of Fig. 1, the results from
Net-Proton, Net-cBaryon and Net-Baryon are almost consistent with
each other, especially in the central collisions. At the same time, the
trends of Kurtosis in three transverse momentum windows seems
similar, it suggests Kurtosis has little dependence on the chosen
transverse momentum window. Together with the previous results of
windows size dependence~\cite{ZHOUY-MPE}, we argue that Kurtosis is
independent on acceptance. We also calculate the transverse momentum window cut effect
on Skewness and find that Skewness is
an observable which does not depend on acceptance.

As an ideal observable for the signatures of CP, it should be only
sensitive to the early stage of the collisions. Therefore, it is
important to investigate the behavior of these observables as a
function of evolution time. In Fig. 2 we study the Kurtosis of
net-proton, net-cBaryon, net-Baryon distributions as a function of
evolution time by UrQMD model. The measured values of Kurtosis of
net-Baryon, net-cBaryon and net-proton distribution keep a constant.
The results suggest that the Kurtosis of the net-baryon distribution
in the final state can reflect the distribution of the early stage.
Most importantly, The system's evolution does not affect the
signature.

Based on the UrQMD model calculation, we find the Kurtosis can provide the information
of the signature of the early stage. With AMPT StringMelting,
we can illustrate different effects on Kurtosis
in time evolution clearly. In Fig. 3 we study the Kurtosis as a function of
centralities in scenario (a)parton, (b)w/o ART , (c)$\mathrm {NTMAX}$ = 3,
(d)$\mathrm {NTMAX}$ = 150 with $0<p_{T}<1$($\mathrm {GeV}/c$) and $|y|<0.5$ at
$\sqrt{s_{NN}}$ =17.3 GeV, respectively. From scenario (a) to (b), the system experiences the hadronization process.
Comparing panel (a) and (b) of Fig. 3, we find the Kurtosis of the distributions
originate from early partonic phase, the quark coalescence-like
hadronization almost doesn't affect  Kurtosis.
\begin{figure}
\resizebox{!}{60mm}{\includegraphics{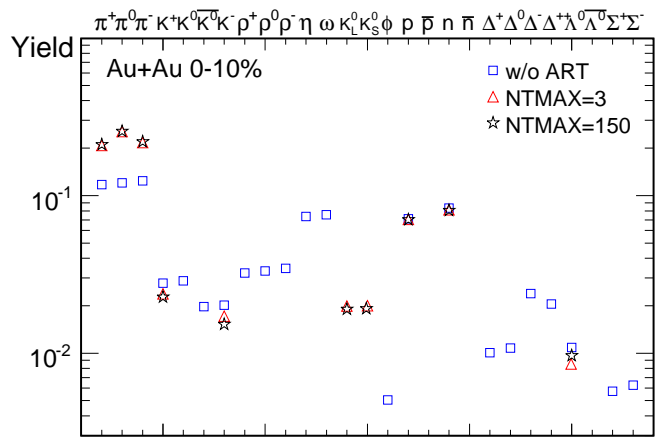}}
\caption{\label{fig:epsart} (Color online) Different particles
yields in time evolution within AMPT StringMelting at
$\sqrt{s_{NN}} $=17.3 GeV.}
\end{figure}
In AMPT StringMelting, the hadron cascade procedure is regarded as two processes,
resonance decay and hadron rescattering. From scenario (b) w/o ART to scenario (c) $\mathrm {NTMAX}$ = 3, the system experiences the resonance decay process.
Fig. 4 is the particles' yield in different evolution time in central collision
at $\sqrt{s_{NN}}$ = 17.3 GeV by AMPT StringMelting. We can find all resonances decay before $t_{\mathrm {cut}}=0.6 \mathrm {fm}/c$
and the certain particle yield doesn't change after $\mathrm {NTMAX}$ = 3.
Here the yield is the ratio of a certain particle number to the total particles number.
When we compare panel (b) and (c) of Fig. 3, it seems
that the resonance decay process doesn't affect Kurtosis.

On the other hand, large hadron rescattering effect may destroy
the fluctuation which originates from the early stage of heavy ion
collision. This effect depends on two factors: one is the time that
particles go through the collision region, the other is the density
in the collision region~\cite{ZQH-PRC}. In AMPT StringMelting, we
can study the hadron rescattering effect on higher moments by
controlling the termination time of hadron cascade($t_{\mathrm
{cut}}$). $\mathrm {NTMAX}$ = 3 is regarded as that there is no hadron
rescattering, while $\mathrm {NTAMX}$ = 150 is
corresponding to termination time of 30 $\mathrm {fm}/c$. From panel
(c) and (d) of Fig. 3, it suggests the observed value of Kurtosis
nearly does not change and almost keeps the same trend, even though
the system experiences full hadronic rescatterings. It is reasonably
to conclude that hadron rescatterings do not have clear influence on
Kurtosis.

\begin{figure}
\resizebox{!}{60mm}{\includegraphics{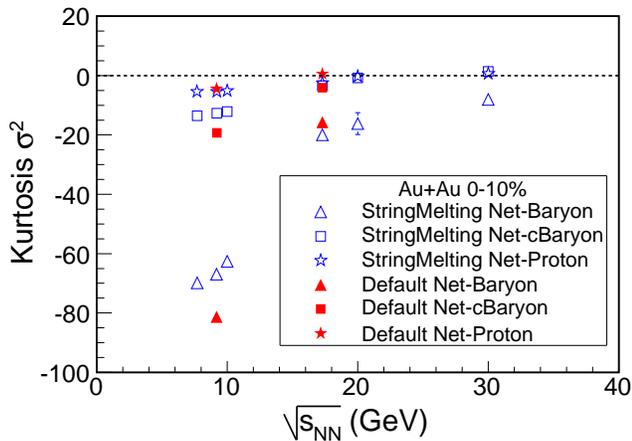}}
\caption{\label{fig:epsart} (Color online) The beam energy
dependence of Kurtosis$\cdot\sigma^{2}$ of net-proton, net-cBaryon,
net-Baryon distributions by AMPT StringMelting.}
\end{figure}

Besides Kurtosis and Skewness, the product of Kurtosis and Variance
(Kurtosis$\cdot\sigma^{2}$) of the net-baryon distribution has also
been used to search for the CP. This observable is related to the
ratio of quartic ($4^{\mathrm {th}}$ order) and quadratic
($2^{\mathrm {nd}}$ order) cumulates of baryon number
susceptibilities. Lattice QCD and some QCD based model~\cite{MC-PRD}
predicted that the fluctuation of Kurtosis$\sigma^{2}$ changed
rapidly at the transition temperature $T_{C}$. Furthermore, it is
predicted that baryon number susceptibilities will diverge when
close to the CP, this will bring about the deviation of
Kurtosis$\cdot\sigma^{2}$ from being constant. All of the works
argue that Kurtosis$\cdot\sigma^{2}$ is a worthful observable.
However, as we presented before~\cite{ZHOUY-MPE}, the
Kurtosis$\cdot\sigma^{2}$ is dependent on the acceptance, namely,
lager acceptance lead to smaller measured value. Also, as shown in
Fig. 5, results of Kurtosis$\cdot\sigma^{2}$ of net-baryon
distribution from two versions of AMPT model have a strong
dependence on the collision energy, it is quite different from
net-proton distribution. This difference may due to baryon stopping
effect in low energy region. The similar result can be found in
Ref.~\cite{TS-arXiv-Keff} by UrQMD calculation.
\\

\section{Summary and outlook}

In this paper we study fluctuation of higher moments of net-baryon
distriubtion with AMPT and UrQMD models. Based on these
results, we find the higher moments of net-baryon distribution
are independent on the chosen transverse momentum window. Together
with the previous results of rapidity and transverse momentum
windows size dependence, we conclude that Kurtosis and Skewness are
independent on acceptance.

We also study the time evolution effects and late hadronic effects
on higher moments of net-baryon distribution in Au+Au collisions at
$\sqrt{s_{NN}}=17.3$ GeV with UrQMD and AMPT model. The results seem
that quark coalescence hadronization process, resonances decay
process and hadronic rescatterings don't affect the trend and value
of Kurtosis and Skewness as a function of centralities. Thus, they
should be good oberservables to searching for the possible critical
point predicted by QCD, non monotonic behaviors of the Kurtosis and
Skewness as a function of beam energy or centrality will demonstrate
the existence of a CP. The Kurtosis$\cdot\sigma^{2}$ of net-proton
distribution show great difference to that of net-baryon
distribution when collisions energy is lower than $\sqrt{s_{NN}}$ =
20 GeV, it deviates from the theoretical predictions.

In the future, the RHIC Beam Energy Scan and GSI
Facility for Antiproton and Ion Research will provide the
possibility to locate the critical point in experiment. Higher moments of
net-baryon distribution should be one of the powerful tools.


\section{Acknowledgments}
We thanks Prof. N. Xu and Prof. C. M. Ko for suggestions, also
thanks Dr. G.L.Ma and J.X.Du for useful discussions on AMPT model. This
work were supported by the National Natural Science Foundation of
China under Grants 10775058, the MOE of China under Grant IRT0624,
the MOST of China under Grant 2008CB817707.

%

%
\end{document}